\documentclass[twocolumn]{aastex62}

\usepackage{graphicx}
\usepackage{natbib}

\received{January 1, 2018}
\revised{January 7, 2018}
\accepted{\today}

\submitjournal{ApJL}

\shorttitle{hcooh in disk}
\shortauthors{Favre et al.}

\begin{document}

%-----------------------------------------------------------------------
% TITLE & AUTHORS
%-----------------------------------------------------------------------
%
\title{First detection of the simplest organic acid in a protoplanetary disk\footnote{This Letter makes use of the following ALMA data: ADS/ JAO.ALMA$\#$2015.1.00845.S (PI C. Favre). ALMA is a partnership of ESO (representing its member states), NSF (USA) and NINS (Japan), together with NRC (Canada), NSC and ASIAA (Taiwan), and KASI (Republic of Korea), in cooperation with the Republic of Chile. The Joint ALMA Observatory is operated by ESO, AUI/NRAO and NAOJ.}}

\correspondingauthor{C\'ecile Favre}
\email{cfavre@arcetri.astro.it}

\author[0000-0002-5789-6931]{C\'ecile Favre}
\affil{INAF-Osservatorio Astrofisico di Arcetri, Largo E. Fermi 5, I-50125, Florence, Italy}

\author{Davide Fedele}
\affiliation{INAF-Osservatorio Astrofisico di Arcetri, Largo E. Fermi 5, I-50125, Florence, Italy}

\author{Dmitry Semenov}
\affiliation{Max Planck Institute for Astronomy, Königstuhl 17, 69117 Heidelberg, Germany}
\affiliation{Department of Chemistry, Ludwig Maximilian University, Butenandtstr. 5-13, D-81377 Munich, Germany}

\author{Sergey Parfenov}
\affiliation{Ural Federal University, 51 Lenin Str., Ekaterinburg 620000, Russia}

\author{Claudio Codella}
\affiliation{INAF-Osservatorio Astrofisico di Arcetri, Largo E. Fermi 5, I-50125, Florence, Italy}

\author{Cecilia Ceccarelli}
\affiliation{Univ. Grenoble Alpes, CNRS, IPAG, F-38000 Grenoble, France}

\author{Edwin~A. Bergin}
\affiliation{Department of Astronomy, University of Michigan, 1085 South University Avenue, Ann Arbor, Michigan 48109, USA}

\author{Edwige Chapillon}
\affiliation{Laboratoire d'astrophysique de Bordeaux, Univ. Bordeaux, CNRS, B18N, allée Geoffroy Saint- Hilaire, 33615 Pessac, France}
\affiliation{IRAM, 300 Rue de la Piscine, F-38046 Saint Martin d'H\`eres, France}

\author{Leonardo Testi}
\affiliation{European Southern Observatory, Karl-Schwarzschild-Str. 2, 85748 Garching, Germany}
\affiliation{Excellence Cluster Universe, Boltzmannstr. 2, 85748 Garching, Germany}
\affiliation{INAF-Osservatorio Astrofisico di Arcetri, Largo E. Fermi 5, I-50125, Florence, Italy}

\author{Franck Hersant}
\affiliation{Laboratoire d'astrophysique de Bordeaux, Univ. Bordeaux, CNRS, B18N, allée Geoffroy Saint- Hilaire, 33615 Pessac, France}

\author{Bertrand Lefloch}
\affiliation{Univ. Grenoble Alpes, CNRS, IPAG, F-38000 Grenoble, France}

\author{Francesco Fontani}
\affiliation{INAF-Osservatorio Astrofisico di Arcetri, Largo E. Fermi 5, I-50125, Florence, Italy}

\author{Geoffrey~A. Blake}
\affiliation{Division of Geological and Planetary Sciences, MC 150-21, California Institute of Technology, 1200 East California Boulevard, Pasadena, California 91125, USA}

\author{L.~Ilsedore Cleeves}
\affiliation{Harvard-Smithsonian Center for Astrophysics, 60 Garden Street, Cambridge, Massachusetts 02138, USA}

\author{Chunhua Qi}
\affiliation{Harvard-Smithsonian Center for Astrophysics, 60 Garden Street, Cambridge, Massachusetts 02138, USA}

\author{Kamber~R. Schwarz}
\affiliation{Department of Astronomy, University of Michigan, 1085 South University Avenue, Ann Arbor, Michigan 48109, USA}

\author{Vianney Taquet}
\affiliation{INAF-Osservatorio Astrofisico di Arcetri, Largo E. Fermi 5, I-50125, Florence, Italy}

%-----------------------------------------------------------------------
% ABSTRACT
%-----------------------------------------------------------------------
%
\begin{abstract}

The formation of asteroids, comets and planets occurs in the interior of protoplanetary disks during the early phase of star formation. Consequently, the chemical composition of the disk might shape the properties of the emerging planetary system. In this context, it is crucial to understand whether and what organic molecules are synthesized in the disk. In this Letter, we report the first detection of formic acid (HCOOH) towards the TW Hydrae protoplanetary disk. The observations of the trans-HCOOH 6$_{(1,6)-5(1,5)}$ transition were carried out at 129~GHz with ALMA. We measured a disk-averaged gas-phase t-HCOOH column density of $\sim$ (2-4)$\times$10$^{12}$~cm$^{-2}$, namely as large as that of methanol. HCOOH is the first organic molecules containing two oxygen atoms detected in a protoplanetary disk, a proof that organic chemistry is very active even though difficult to observe in these objects. Specifically, this simplest acid stands as the basis for synthesis of more complex carboxylic acids used by life on Earth.

\end{abstract}

%-----------------------------------------------------------------------
% KEYWORDS
%-----------------------------------------------------------------------
%

%%%%%%%%%%%%%%%%%%%%%%%%%%%%%%%%%%%%%%%%%%%%%%%%%%%%%%%%%%%%%%%%%%%%%%%%%%%%%%%%

%===============================================================
%===============================================================

%-----------------------------------------------------------------------
%----------INTRODUCTION -------------
%-----------------------------------------------------------------------
\section{Introduction} \label{sec:intro}
Life on Earth is based on different combinations of a relatively small number of key organic constituents, synthesized from simpler building-blocks including amino acids, phosphates, esters, organic acids, sugars and alcohols. Some of these prebiotic molecules have been discovered in Solar-type star forming regions \citep{Cazaux:2003,Jorgensen:2012,Kahane:2013} as well as in meteoritic \citep{Kvenvolden:1970} and cometary \citep{Elsila:2010,Altwegge:2016} material. Astronomers have long wondered whether the organic chemistry during the star and planet formation process is inherited by planets and small bodies of the final planetary system and what the key organic molecules are. In order to establish this missing link, it is mandatory to understand how organic chemistry evolves during the protoplanetary disk phase, the last step before the planet formation. So far, however, only about twenty molecules have been detected in protoplanetary disks \citep{Dutrey:2014}. Among them are small hydrocarbons, such as c-C$_3$H$_2$ \citep{Qi:2013d,Bergin:2016}, and cyanides, HC$_3$N and CH$_3$CN {\citep{Chapillon:2012,Oberg:2015,Bergner:2018}}, as well as two organic O-bearing molecules, methanol, CH$_3$OH \citep{Walsh:2016}, and formaldehyde, H$_2$CO { \citep{Qi:2013a,Loomis:2015,Oberg:2017,Carney:2017}}, believed to be the first step towards a complex organic chemistry. The search for relatively large molecules in protoplanetary disks remains challenging \citep{Walsh:2014}: high sensitivity and resolution (spatial and spectral) are required. The unprecedented sensitivity of the Atacama Large Millimetre/Submillimetre Array (ALMA) makes this instrument the most suitable for such a study.

In this Letter, we { focus on formic acid (HCOOH), a key organic molecule as the carboxyl group (C($=$O)OH) is one of the main functional groups of amino acids (the structural units of proteins). Indeed, this species is involved in a chemical route leading to glycine, the simplest amino acid \citep[e.g., see][]{Basiuk:2001,Redondo:2015}.
HCOOH is unambiguously detected toward both high- and low-mass star forming regions \citep[e.g., see][]{Liu:2002,Lefloch:2017}. Here, we report the first detection of HCOOH with ALMA} towards the protoplanetary disk surrounding the closest { \citep[59~pc,][]{Gaia-Collaboration:2016}} Solar-type young star TW Hya. TW Hya is a relatively old (3-10~Myr) T Tauri star of about 0.7 M$_{\odot}$, which is surrounded by a gas-rich disk, whose mass is $\ge$ 0.006~M$_{\odot}$ \citep{Bergin:2013,Trapman:2017} and which shows prominent rings and gaps in gas and dust emission, perhaps a signature of ongoing planet formation. Notably, combined HD and CO observations towards TW Hya suggest that a substantial fraction of carbon reservoir is absent from the gas phase and could be stored in organic species, possibly in relatively complex ones {\citep{Favre:2013,Schwarz:2016,Zhang:2017}}. 
This detection of new class of organic species has important consequences since a relatively high, observationally-derived, abundance of HCOOH along with that CH$_3$OH \citep[$\chi$ $\sim$ 3$\times$10$^{-12}$--4$\times$10$^{-11}$, see ][]{Walsh:2016}, likely imply a rich organic chemistry in protoplanetary disks, at the epoch of planet formation. 

In Sections~\ref{sec:obs}, we describe our observations. {Results and analysis are given in Sect.~\ref{sec:results}. Physico-chemical modeling and further discussion are presented in Sect.~\ref{sec:discussion}. }

%===============================================================

%-----------------------------------------------------------------
%-------- OBSERVATIONS-------------------
%-----------------------------------------------------------------
\section{Observations and data reduction} \label{sec:obs}

In this Letter, we focus on the trans-HCOOH 6$_{(1,6)-5(1,5)}$ transition (i.e the OH functional group is on the opposing side to the single C-H bond) emitting at 129671.82~MHz. 
{We used the spectroscopic data parameters from \citet{Kuze:1982} that are available at the Cologne Database for Molecular Spectroscopy line catalog \citep[CDMS,][]{Muller:2005}.}
This transition has an appropriate low upper state energy level ($\sim$25~K) and high line strength ($\sim$12~D$^{2}$), making it ideal for searching for this molecule in the gas with a temperature range of about 10-40~K, which is typical for the outer regions of protoplanetary disks surrounding young Sun-like stars like TW~Hya.
{In addition, we note that 6 lines from the cis conformer, c-HCOOH, at 131~GHz were also covered by the observations. However, the energy barrier to internal trans conversion to cis is really high \citep[$\sim$1365~cm$^{-1}$, see][]{Winnewisser:2002} making their detection unlikely.}  

\begin{figure}
\centering
\includegraphics[trim={0 4cm 0 0},angle=0,width=8cm]{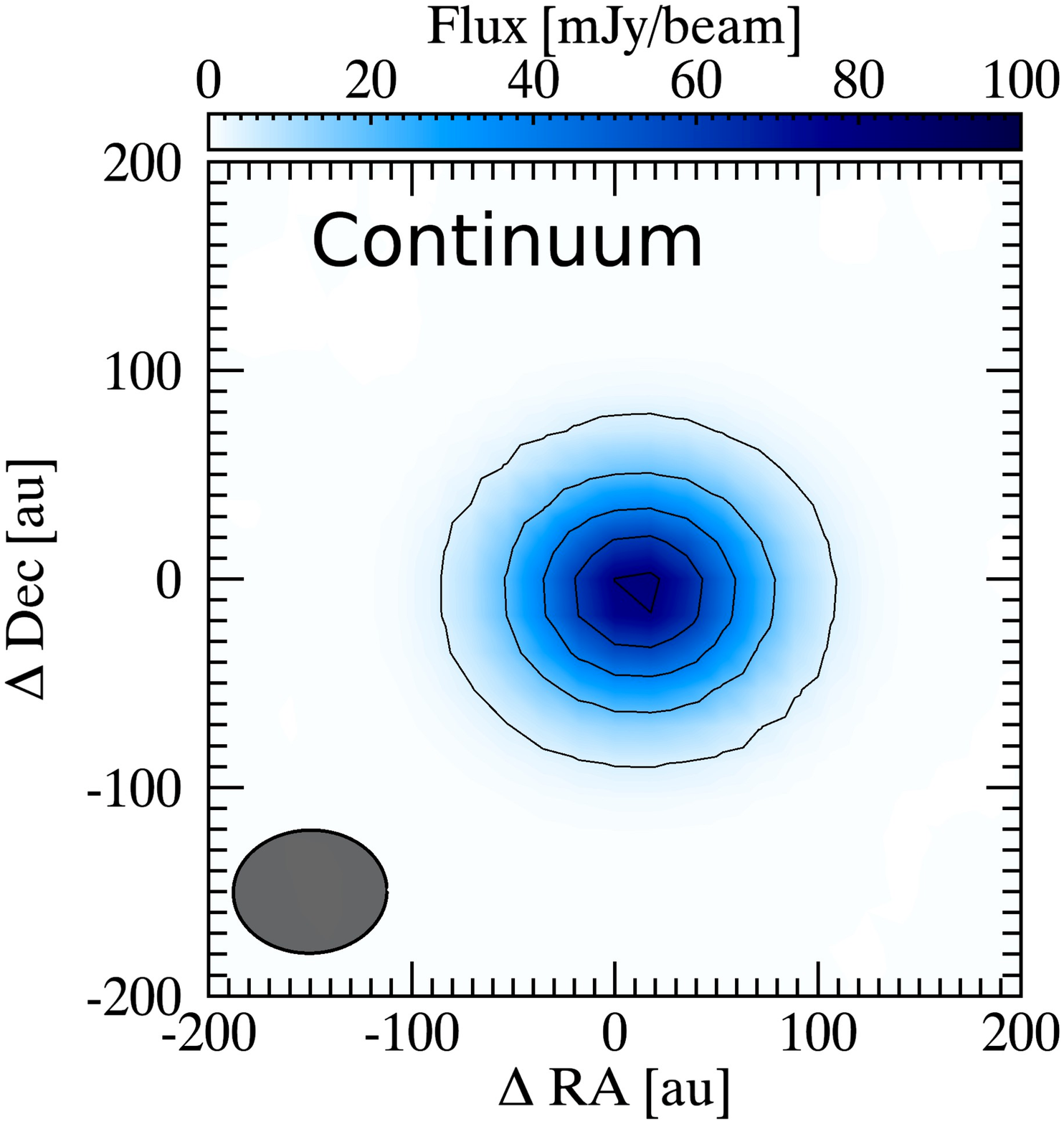}
\caption{129~GHz continuum emission as observed with ALMA towards TW Hya. The contours and level step are at 14$\sigma$ (where 1$\sigma$ = 1.4~mJy~beam$^{-1}$).}
\label{fg1}
\end{figure}

{The observations were performed toward TW~Hya} on 2016 May 31 and on 2016 June 1 with 39 antennas and baselines from 15.1~m {(6.5~k$\lambda$)} up to 713.1~m {(310.0~k$\lambda$)} with an on source time of about 39~min with ALMA, during Cycle 3.
The phase-tracking center was: $\alpha_{J2000}$ = 11$^{h}$01$^{m}$51\fs875, $\delta_{J2000}$ =$-$34$\degr$42$\arcmin$17$\farcs$155. The spectral setup consisted into: (i) six spectral windows each of 960 channels with a {channel width used during the observations} of 244.141 kHz (about 0.6 km~s$^{-1}$) covering about 1.4~GHz between 129.538~GHz and 132.405~GHz and, (ii) one spectral window, centred at 130.973~GHz, of 2~GHz bandwidth (128 channels each of 15.625~MHz) {for continuum imaging}. The quasars J1037-2934 and J1103-3251 were used as calibrators for bandpass and phase for these observations. Ganymede and Titan were used as flux calibrators for the observations performed in May and June, respectively.
Data reduction and continuum subtraction were performed through the version 4.5.3 of the Common Astronomy Software Applications, CASA. The continuum emission at 129~GHz was {self-calibrated} with the solutions (gain tables) {applied} to the molecular data. In addition, and still in order to {optimise} the sensitivity, the data were cleaned using a $``$Natural" weighting. The resulting synthesized beam sizes are 1.28$\arcsec$ $\times$ 1.00$\arcsec$ at a P.A.$\sim$ $-$89$\degr$ and 1.29$\arcsec$ $\times$ 1.02$\arcsec$ at a P.A. of about 90$\degr$ for the continuum (see Fig.~\ref{fg1}) and line images, respectively.

%------------------------------------------------------------------
% --- FIGURE 1 ---
%-----------------------------------------------------------------

%===============================================================

%
%-----------------------------------------------------------------------
%-------------------------------------------- RESULTS ------------------
%-----------------------------------------------------------------------
\section{Results and analysis}
\label{sec:results}

% *** t-HCOOH ***
\subsection{t-HCOOH}

\begin{figure*}
\centering
\includegraphics[angle=0,width=8cm]{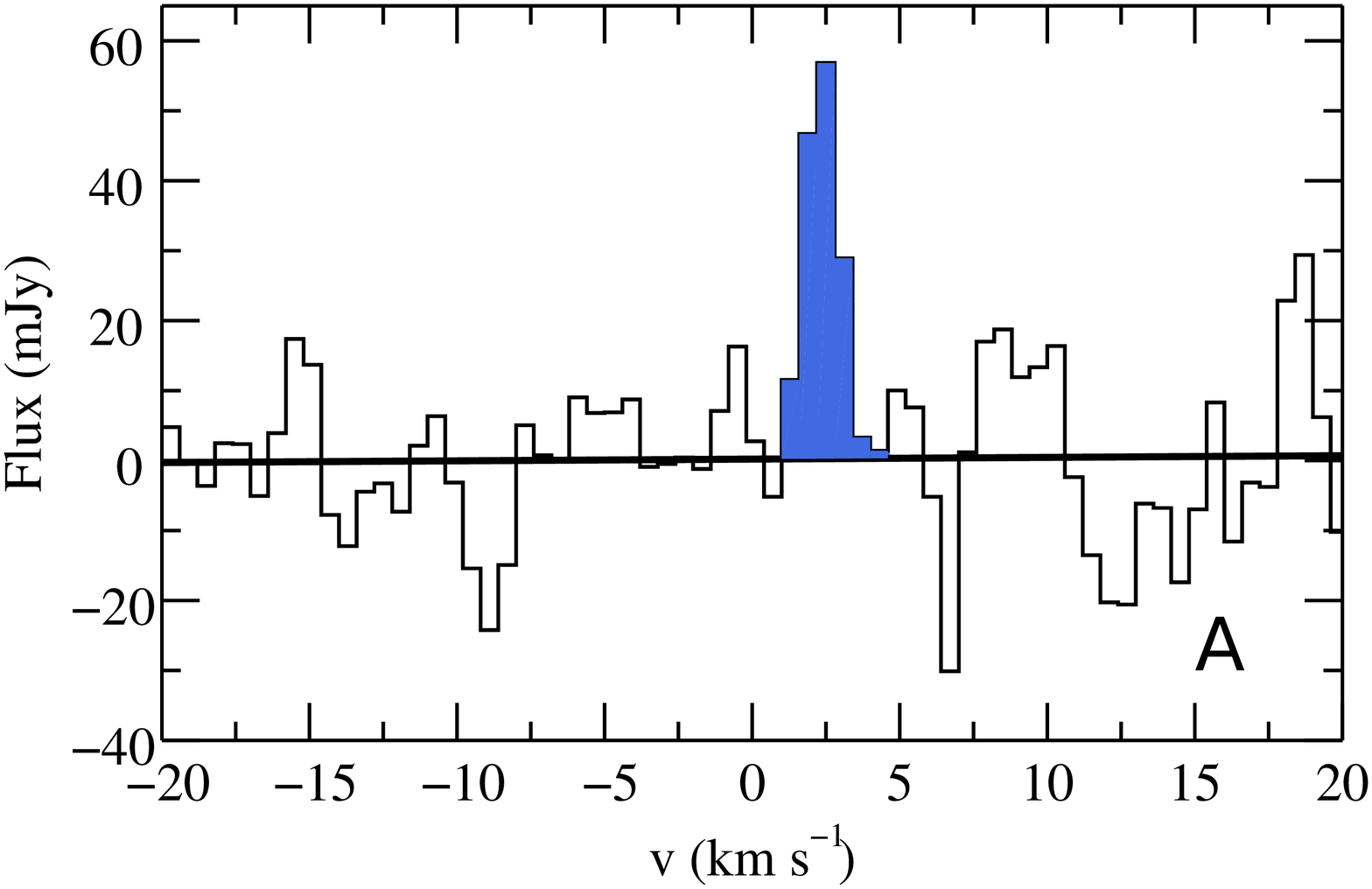}
\includegraphics[angle=0,width=8cm]{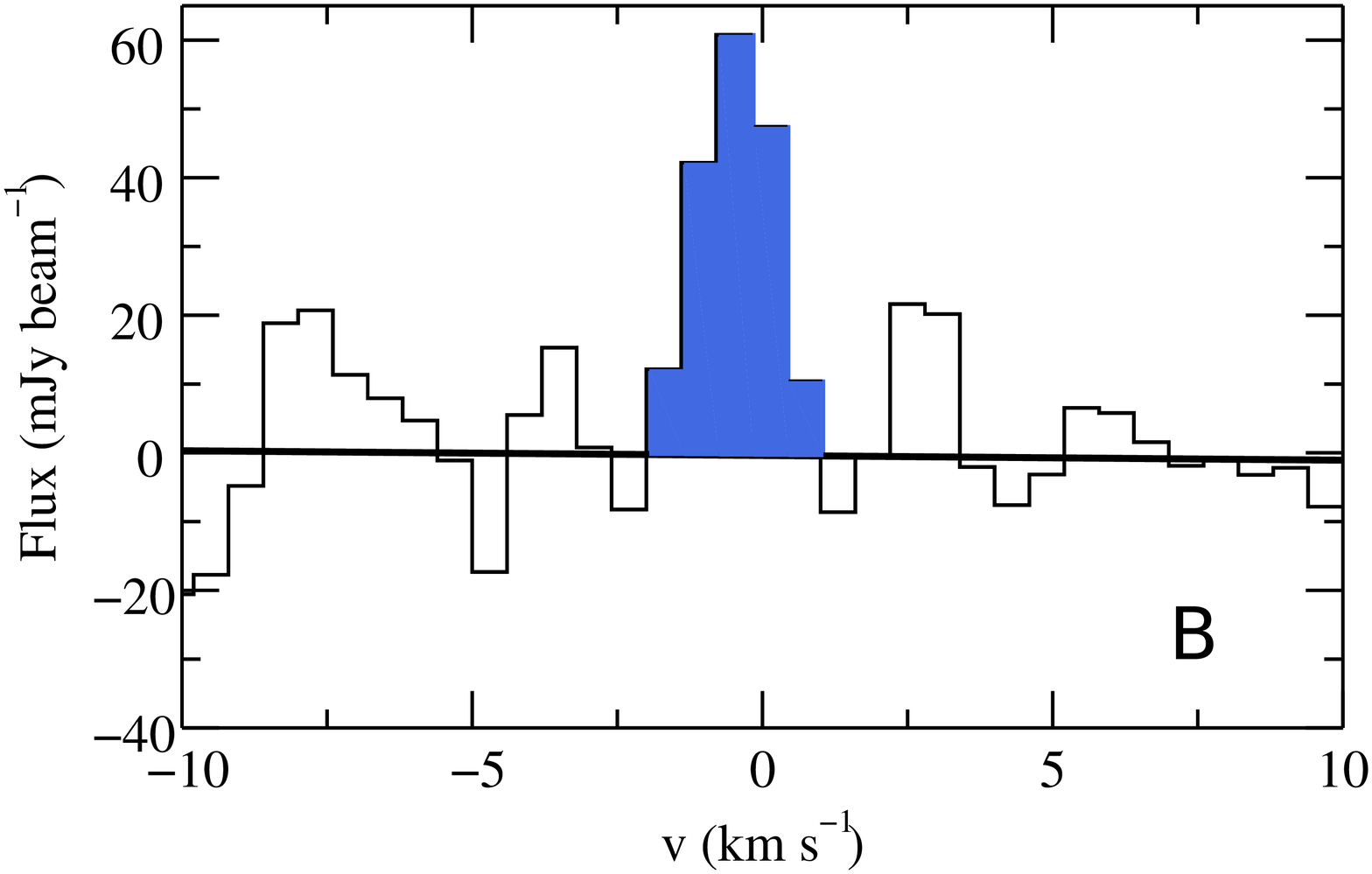}
\caption{ {{\it Panel a}: TW~Hya disk-averaged HCOOH spectrum extracted from the original datacube within 400~au . {\it Panel b}: TW~Hya HCOOH spectrum resulting from the pixel spectral stacking method by \citet{Yen:2016}. The signal to noise ratio is $\sim$5.}}
\label{fg2}
\end{figure*}

We detected the 129 GHz HCOOH line with a signal-to-noise ratio of about 4. 
{Figure~\ref{fg2} shows the disk-averaged spectrum extracted from the native dataset using an elliptical aperture of about 400~au {to take into account uncertainties on the position and the spatial extent of the HCOOH emission.} Fig.~\ref{fg2} also displays the HCOOH spectrum resulting from the pixel spectral stacking method \citep[assuming a Keplerian disk, see][]{Yen:2016}, which independently strengthens our detection.} Even though we have only one detected line, we can firmly confirm the identification of HCOOH because this is  \textit{i)} the brightest line expected in the observed range, \textit{ii)} the same line was clearly detected in the Solar-like regions L1157--B1 and NGC1333--IRAS4A with the IRAM-30m telescope as part of the Astrochemical Surveys At IRAM large program {\citep[see][]{Lefloch:2017,Lefloch:2018} and, \textit{iii)} two independent methods show a detection above the 4$\sigma$ level (see Fig.~\ref{fg2}).}

To further enhance the spatial signal-to-noise ratio, sophisticated data processing has been performed (see Section~\ref{kepmask}). 

%
%------------------------------------------------------------------
% --- FIGURE 2 ---
%-----------------------------------------------------------------

%
%------------------------------------------------------------------
% --- FIGURE 3 ---
%-----------------------------------------------------------------
\begin{figure*}
\centering
\includegraphics[trim={0 4cm 0 0},clip,angle=0,width=8cm]{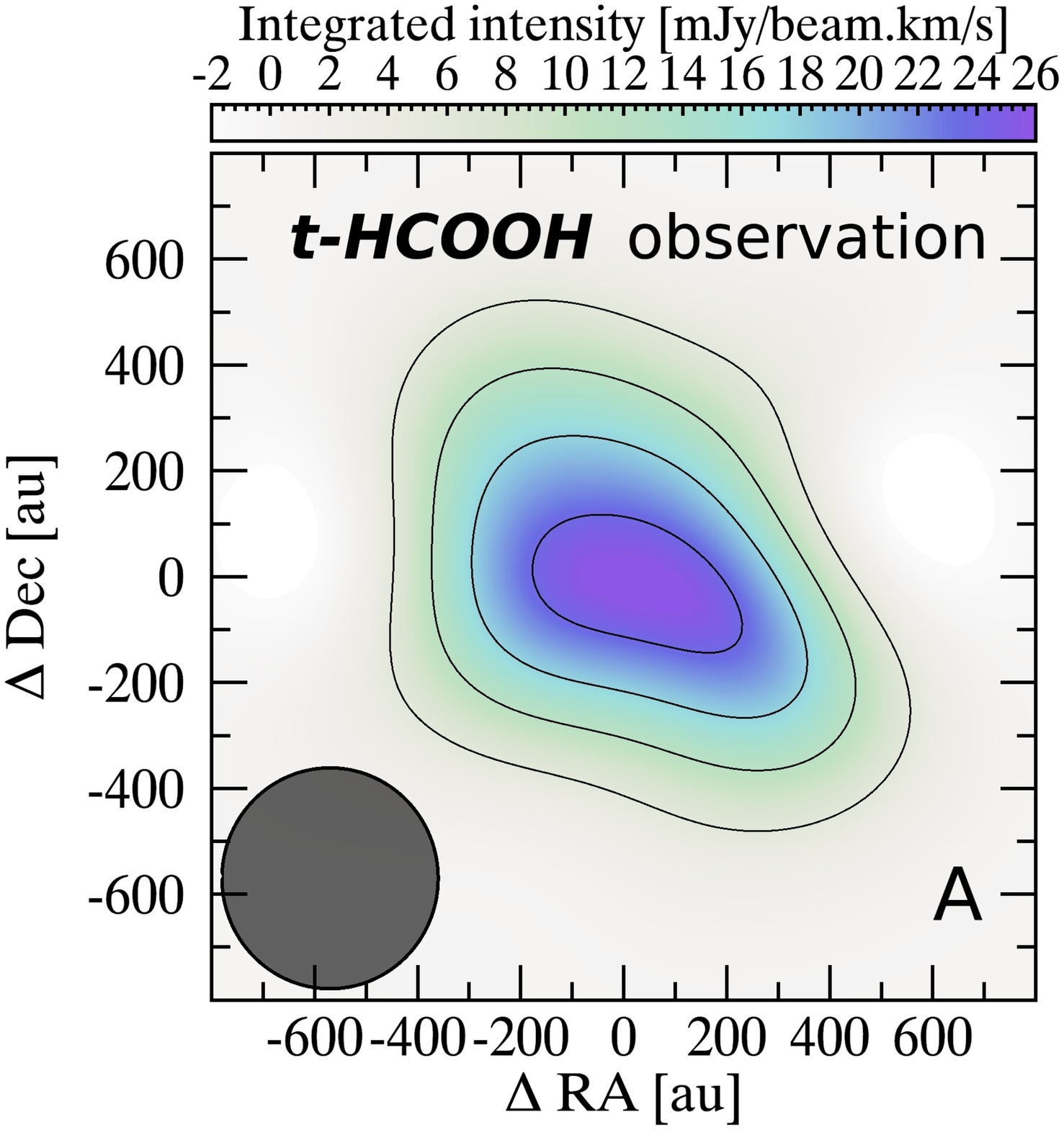}
\includegraphics[trim={0 4cm 0 0},clip,angle=0,width=8cm]{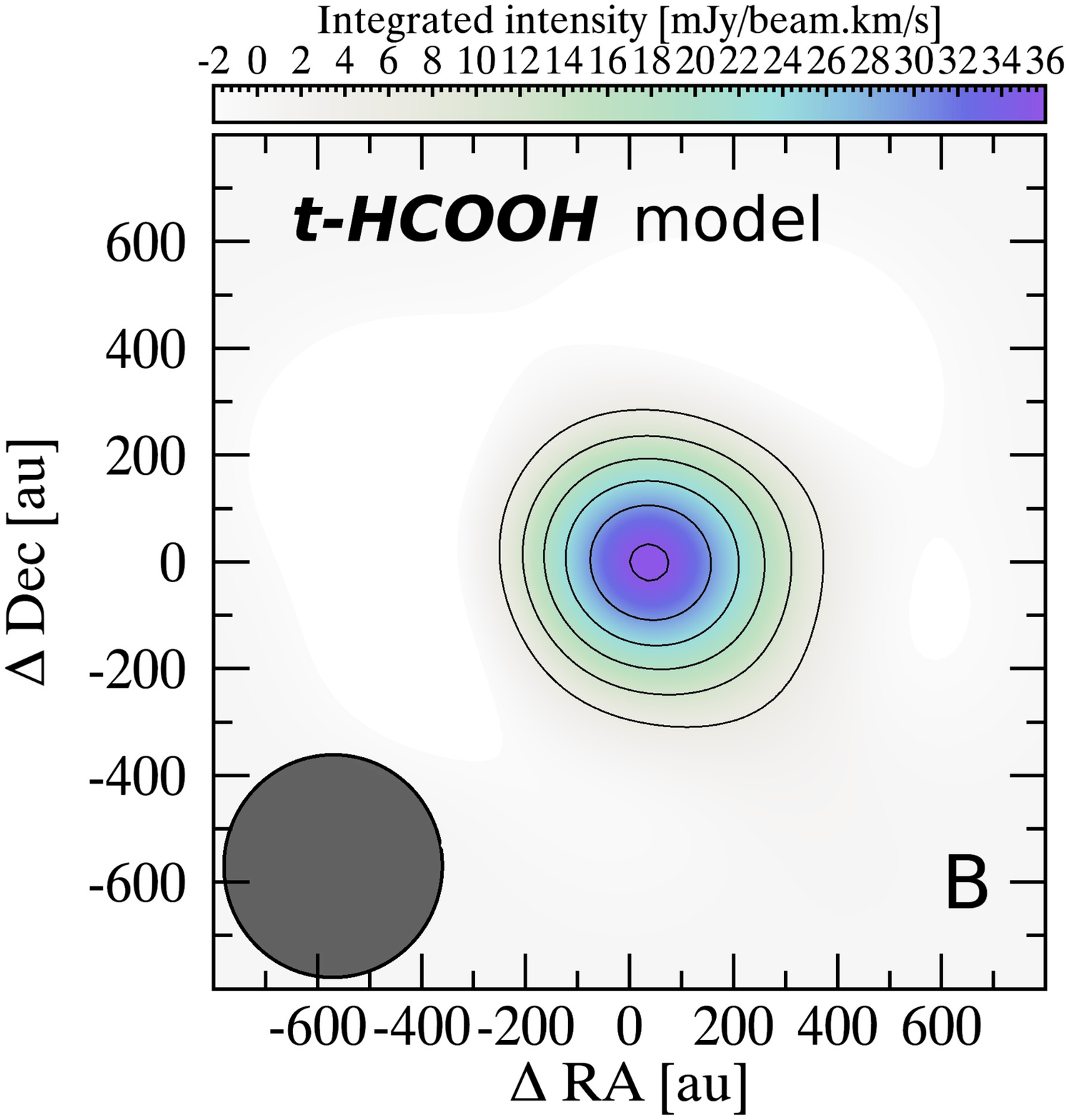}
\caption{{\it Panel a}: Observed gas-phase t-HCOOH emission integrated over the line profile {after applying a Keplerian mask} and smoothed at resolution equivalent to the TW~Hya disk size. The contours and level step are at 3$\sigma$ (where 1$\sigma$ = 2~mJy~beam$^{-1}$~km~s$^{-1}$, {in the smoothed and masked data}). {\it Panel b}: Same as {\it A} but model. The synthesized beam is shown in the bottom left corner of panels A and B.}
\label{fg3}
\end{figure*}

% *** Keplerian mask ***
\subsection{Data processing: keplerian mask}\label{kepmask}
Assuming that the disk is in Keplerian rotation, one can significantly improve the signal- to-noise by the use of a Keplerian mask \citep{Salinas:2017a,Carney:2017}. More specifically, the latter selects the regions in the data where the signal is expected to emit, following a disk velocity pattern (including the rotational and the systemic velocities of the object) and a disk size.

The parameters we used to define out TW Hya Keplerian mask for the t-HCOOH emission are as follows:

\begin{itemize}  
	  \setlength\itemsep{-0.3em}
\item[-] stellar mass of 0.7~M$_{\odot}$,
\item[-] disk inclination of 7$\degr$,
\item[-] disk position angle of 155$\degr$,
\item[-] systemic velocity of 2.7~km~s$^{-1}$,
\item[-] distance of 59 pc, consistent with the recent GAIA measurements \citep{Gaia-Collaboration:2016},
\item[-] outer radius of 400~au {(to be consistent with the data).}
\end{itemize}
The consistency of the mask we used in this study has been checked on the c-C$_3$H$_2$ (3-2) line emission at 145089.6~MHz which is clearly observed above the 12$\sigma$ level as part of our project (Favre et al. in prep).
After applying the above Keplerian mask to the HCOOH data, {the signal to noise ratio was enhanced by a factor $\sim$5. Note that to compute the "bootstrapped" noise, the same keplerian mask was applied to channels with no signal that were far away from the ones in which HCOOH is emitting \citep[see][]{Bergner:2018}.}
Then, the resulting t-HCOOH emission was integrated over the line profile, from v=0.9 to 3.4~km~s$^{-1}$ and smoothed by a Gaussian kernel of 7$\arcsec$ $\times$ 7$\arcsec$ to spatially enhance the signal to noise ratio. Figure~\ref{fg3} displays the resulting {smoothed} disk averaged t-HCOOH spatial distribution (with a detection at {a peak to noise ratio of 12$\sigma$), after applying the Keplerian mask}.

It is important to note that the low spectral and spatial resolution of our observations does not allow us to disentangle the spatial structure of the emission nor to investigate the kinematics: \textit{i)} the line peak is slightly displaced with respect to the LSR velocity of the source (2.86~km~s$^{-1}$) by at most one channel, but this displacement is consistent with the spectral resolution (0.6~km~s$^{-1}$), \textit{ii)} the low angular resolution observations along with the convolution procedure, lead to at least 100~au of uncertainty regarding the spatial extent of HCOOH.

% *** N_hcooh ***
\subsection{Column densities and relative abundances}\label{coldensab}

The measured disk averaged line integrated intensity over the line profile, F = (89$\pm$12)$\times$10$^{-3}$~Jy~km~s$^{-1}$, corresponds {to a t-HCOOH source averaged column density of (2.9$\pm$1.3)$\times$10$^{12}$~cm$^{-2}$ and (3.5$\pm$0.8)$\times$10$^{12}$~cm$^{-2}$ for an excitation temperature of 10 and 40~K, respectively. The 10-40~K temperature range is the typical one} where the molecular gas is expected to be emissive. Our observations suggest that HCOOH emission appears centrally peaked with extension beyond 200~au. Although our present low-resolution observations do not allow us to constrain where exactly formic acid is emitting within the TW Hya disk, one could simply assume that all organic O-bearing molecules emit within the same region {if they share grain-surface formation chemistry}. As methanol was previously detected towards TW Hya \citep{Walsh:2016}, one can estimate the fraction of formic acid with respect to methanol, often believed to be a starting molecule from which more complex organics are synthesized either in the gas-phase \citep{Charnley:1992,Balucani:2015} or within the icy mantles of dust grains \citep{Garrod:2006,Semenov:2011}. In that light, we obtain an average abundance t-HCOOH/CH$_3$OH ratio lower than and/or equal to 1, which is about one order of magnitude higher than the ratio measured in comets \citep{Biver:2014}. Here, we stress that this ratio likely suffers from the fact that we are not sampling the same spatial scales. Indeed, current methanol observations \citep{Walsh:2016} seem to indicate a more compact spatial distribution than our t-HCOOH observations. Aside from the intrinsic limitations of our observations and that of methanol, this difference in spatial distribution can originate from the different formation pathways.

%===============================================================

%
%----------------------------------------------------------------
%------------Discussion---------
%----------------------------------------------------------------
\section{On the production of HCOOH}
\label{sec:discussion}

From a chemical point of view, the chemistry leading to formic acid is more complex than that responsible for the methanol formation. Indeed, the latter is mostly the result of the hydrogenation of carbon monoxide on the surface of grains, forming H$_2$CO and CH$_3$OH from CO, which is facilitated at low temperatures (about 20~K), which allow \textit{i)} CO molecules to freeze out onto the icy grain surfaces, \textit{ii)} hydrogen diffusion at the surface of grain mantles (prior to their evaporation), \textit{iii)} and consequently, hydrogen to tunnel through energy barriers which would be otherwise insurmountable \citep{Watanabe:2002,Rimola:2014}. Formic acid, on the contrary, cannot be a simple H-- addition process and has to be the result of reaction(s) involving polyatomic species, either on the grain surfaces or in the gas phase \citep{Ioppolo:2011,Skouteris:2018}.

%
%-----------------------------------------------------------------------
%-----------------MODELING ------------------
%-----------------------------------------------------------------------
\subsection{Modeling}
To better understand HCOOH chemistry across the disk, we used a state-of-the-art modeling suite \citep{Parfenov:2017}, that includes a disk physical structure, radiative transfer and chemical modeling that adequately describes the methanol observations in TW Hya disk \citep{Walsh:2016} {and within a factor 5 that of CH$_3$CN by \citet{Loomis:2018}.}

%------------------------------------------------------------------
% --- FIGURE 4 ---
%-----------------------------------------------------------------
\begin{figure*}
\centering
\includegraphics[trim={0 0 4cm 0},clip,angle=270,width=\hsize]{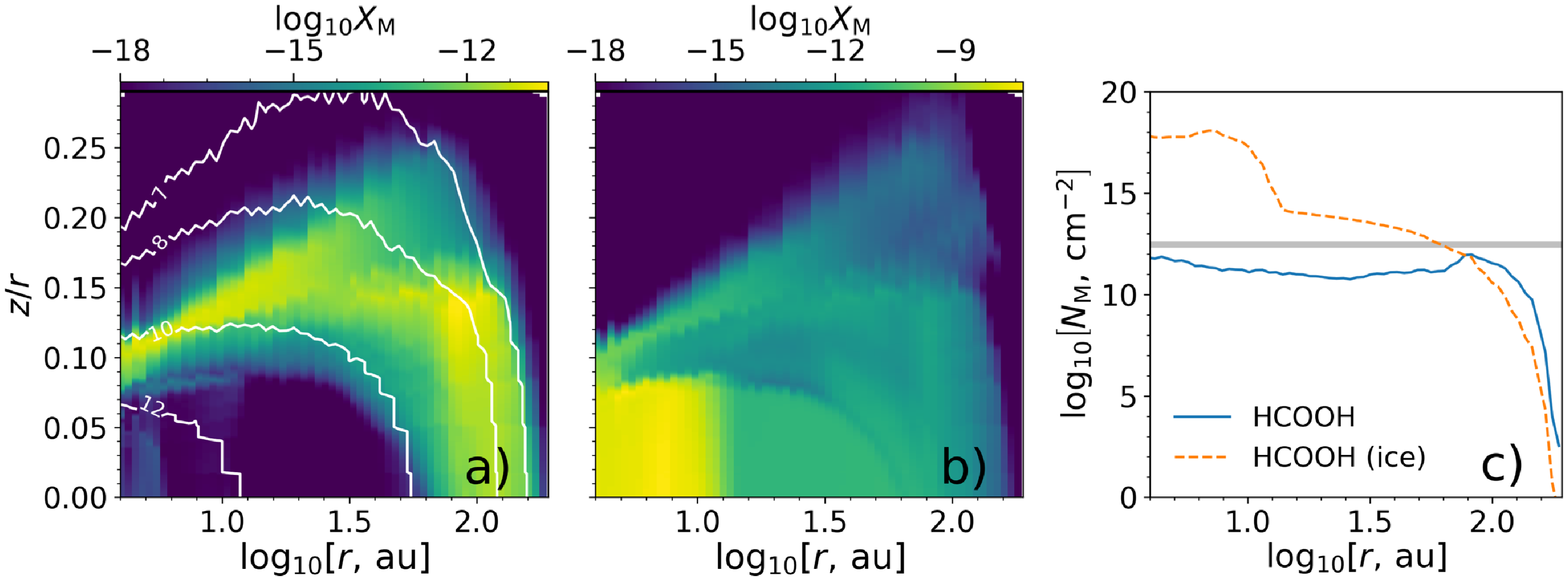}
\caption{Theoretical t-HCOOH relative abundance to molecular hydrogen. {\it Panel a}: Spatial distribution of the gas phase abundance as a function of the radial distance from the central star and the altitude over the midplane. {The white contours denote log10(n$\rm_{H_2}$) in cm$^{-3}$.} {\it Panel b}: Same as A but for the HCOOH ice. {\it Panel c}: Radial distribution of column densities of gas (blue line) and solid ({dashed orange} line) HCOOH. The grey line shows the measured HCOOH column density. The predictions are based on our physical and chemical model \citep[][{see Section~\ref{sec:discussion}}]{Parfenov:2017}.}
\label{fg4}
\end{figure*}

% *** Physical and chemical model ***
\subsubsection{Physical and chemical model of TW Hya disk}\label{phychemmodel}

Our disk physical structure is computed using a thermochemical model by \citet{Gorti2011} for TW Hya. The disk spans 3.9--200~au radially and has separately computed gas and dust temperatures. An accretion rate of 10$^{-9}$~M$_{\odot}$~yr$^{-1}$ and a constant gas-to-dust mass ratio of 100 are assumed.
The adopted parameters of the central star and the incident ultraviolet (UV) and X-ray radiation are representative of those observed in TW Hya. The star has a mass of 0.7~M$_{\odot}$, a radius of 1~R$_{\odot}$, and an effective temperature of 4200~K. A far-UV spectrum with a total far-UV luminosity of 3$\times$10$^{31}$~erg~s$^{-1}$ was used {\citep{Herczeg:2002,Herczeg:2004,Cleeves:2014}}. The adopted X-ray spectrum, covering 0.1-10~keV, has a total X-ray luminosity of 1.6$\times$10$^{30}$~erg~s$^{-1}$.

The chemical structure of the disk was computed with the public gas-grain ALCHEMIC code \citep[see \url{http://www.mpia.de/homes/semenov/disk_chemistry_OSU08ggs_UV.zip},][]{Semenov:2010}. The chemical network is based on the osu.2007 ratefile with recent updates to the reaction rates from Kinetic Database for Astrochemistry \citep[KIDA,][]{Wakelam:2012}, and a high-temperature network \citep{Harada:2010,Semenov:2011}.

{A standard cosmic ray (CR) ionization rate was assumed to be $\zeta_{CR}$ = 1.3$\times$10$^{-17}$~s$^{-1}$, as it does not affect significantly the chemistry in comparison to the stellar X-rays.}
The FUV penetration into the disk is computed in a 1+1D approximation \citep{Semenov:2011} using the visual extinction A$_V$ in the direction towards the central star for the stellar FUV component, and the visual extinction A$\rm^{IS}$ in the vertical direction for the interstellar (IS) FUV component. To compute the X-ray ionization rate, we used the parametrization given by \citet{Armitage:2007} for the average X-ray photon energy of 3~keV. The self-shielding of H$_2$ from photodissociation is calculated following the parametrization given in \citep{Draine:1996}. The shielding of CO by dust grains, H$_2$, and the CO self-shielding are calculated using a precomputed table \citep[see][]{Lee:1996}.

The gas-grain interactions include sticking of neutral species and electrons to dust grains with 100$\%$ probability and desorption of ices by thermal, CRP, and UV-driven processes. We do not allow H$_2$ to stick to grains. The uniformly-sized, compact amorphous silicate particles with the radius a$_d$ = 7 $\micron$ are considered. This grain size represents grain surface per unit gas volume of the size ensemble used in the physical model. A low UV-photodesorption yield of 10$^{-5}$ is adopted for all ices, based on the recent measurements \citep{Cruz_Diaz2016,Bertin2016}. Photodissociation processes of solid species are also taken into account \citep{Garrod:2006,Semenov:2011}. A 1$\%$ probability for chemical desorption is assumed \citep{Garrod:2007}. Surface recombination proceeds solely via Langmuir-Hinshelwood mechanism \citep{Hasegawa:1992}, described by the standard rate equation approach. The ratio of diffusion to binding energy of surface species $ E_{diff}/E_{b}$ is 0.4, consistent with the recent laboratory studies \citep{Cuppen2017}.
A ``low metals'', mainly atomic set of initial abundances \citep{Lee:1998,Semenov:2010} was used (see Table~\ref{tab1}). The adopted thermochemical disk physical model coupled with alchemic code was used to model the entire TW Hya disk chemical evolution over t=1 Myr.

\begin{table}
\centering
\caption{Initial chemical abundances}{\label{tab1}}
\begin{tabular}{cccc}
\hline
\hline
Species & Abundance &  Species & Abundance \\
\hline
ortho-H${_2}$ & 3.75 (-1) & S &9.14 (-8)\\
para-H$_{2}$ & 1.13 (-1) &Si & 9.74 (-9)\\
He & 9.75 (-2) & Fe & 2.74 (-9) \\
O& 1.80 (-4) & Na&2.25 (-9)\\
C &7.86 (-5) & Mg& 1.09 (-8)\\
N & 2.47 (-5) &Cl &1.00 (-9)\\
HD & 1.55 (-5) & P & 2.16 (-10)\\
\hline
\end{tabular}
\end{table}

% *** Radiative transfer ***
\subsubsection{Radiative transfer modeling}\label{radtrans}

To predict the HCOOH line emission from our disk model we used the three-dimensional Monte-Carlo code Line Modeling Engine \citep[LIME,][]{Brinch:2010} with a setup similar to the one used by \citet{Parfenov:2017} to reproduce methanol emission towards TW Hya by \citet{Walsh:2016}.
The radiative transfer calculations were performed assuming local thermodynamic equilibrium {(LTE). Non-LTE calculations could not be performed as the collisional rate coefficients for HCOOH are not available. Nonetheless, as shown in Fig.~\ref{fg4} (see {\it Panel a}), HCOOH is abundant in the disk regions where the H$_2$ density is relatively high (n$\rm_{H2}$ $\ge$ 10$^{7}$~cm$^{-3}$) and 
assuming a ''standard" collisional coefficient of 10$^{-11}$-10$^{-10}$~cm$^{-3}$~s$^{-1}$,
one might expect the line to be thermalized.}

 In order to simulate the spectral averaging within the ALMA channels, the synthetic image cubes produced by LIME (with a spectral resolution of 0.06~km~s$^{-1}$) were averaged along the spectral axis down to the resolution of 0.6~km~s$^{-1}$, which is the same as our observed data. These datacubes were then converted into visibilities using the simobserve task from the CASA package, using the same antenna positions as in the observations. Then noise was added to the visibilities using sm.setnoise and sm.setgain CASA tasks. Finally, the data were deconvolved and the image reconstructed via the clean task. The synthetic beam with the size of 1.31$\arcsec$ $\times$ 0.98$\arcsec$ at a position angle PA of -79.6$\degr$ which well matches the observed beam. The rms noise level in the synthetic disk images is of about 1~mJy~beam$^{-1}$ per channel that is consistent with the observations {at the original spatial resolution.}

%===============================================================

\subsection{{Chemical desorption as the source of HCOOH in disks?}}

{With the above} model we are able to compute the evolution of the formic acid abundance over the age of TW Hya and thus, to reconstruct its spatial distribution within the TW Hya protoplanetary disk (Fig.~\ref{fg3}b). Figure~\ref{fg4} shows the resulting modeled distribution of the HCOOH relative abundance to molecular hydrogen. Our physico-chemical modeling reproduces well the observed column densities within a factor 3 that is commensurate with the uncertainties (from both the observations and the model). In this model, HCOOH is assumed to be formed by the recombination of HCO and OH radicals on the grain surfaces \citep{Garrod:2008a}, and a fraction of it (1$\%$) to be injected into the gas phase {through reactive desorption,} because of the energy released by the reaction itself. In addition, some HCOOH molecules can be released from the ice mantles through cosmic ray particles (CRP) heating of grains, CRP-induced UV-photodesorption, and direct UV desorption in the more irradiated outer disk region.

Interestingly enough, \citet{Skouteris:2018} have recently shown that a new gas-phase scheme of reactions, involving ethanol as a parent molecule, can lead to the formation of formic acid. Using a preliminary 0D-model for disk-like conditions \citep{Semenov:2010}, this new set of reactions does not seem to dominate the production of formic acid in disks. Nevertheless, this leads one to consider other gas phase and/or grain surface processes that may be worth studying in the future as they could contribute to the observed distribution of formic acid.

%
%===============================================================
%
%-----------------------------------------------------------------
%---------Conclusions-----
%-----------------------------------------------------------------
%
\section{Conclusions}

In summary, we report the first detection of the simplest acid, HCOOH, in the protoplanetary disk surrounding a Sun-like young star, TW Hya. Our finding implies that a rich organic chemistry, that can lead to larger organic molecules, likely takes place at the verge of planet formation in protoplanetary disks. Indeed, HCOOH together with methanol and formaldehyde \citep{Walsh:2014} are the most abundant complex molecules detected in protoplanetary disks so far. In the context of an Interstellar-Earth connection, this shows that at least some of the bricks of prebiotic chemistry are already present in a protoplanetary disk expected to be similar to the Solar Nebula that formed our Solar System. Incidentally, our study strengthens the fact that observations of larger complex molecules in this environment remain challenging, even with ALMA with its unprecedented sensitivity. Finally, further improvements in our understanding of the formation of organic molecules, through both laboratory experiments and theory, will help us to target future exploration of the chemical richness of protoplanetary disks, in particular to better characterize where these molecules are expected to be located within the disk.

%
%===============================================================
%
%-----------------------------------------------------------------
%---------ACKNOWLEDGEMENTS-----
%-----------------------------------------------------------------
%
%\section{Acknowledgements}
\acknowledgments
We are very grateful to Nathalie Brouillet for her comments on formic acid in the ISM. This work was supported by \textit{(i)} the Italian Ministry of Education, Universities and Research, through the grant project SIR (RBSI14ZRHR), \textit{(ii)} by the Italian Ministero dell'Istruzione, Università e Ricerca through the grant Progetti Premiali 2012 - iALMA (CUP C52I13000140001), \textit{(iii)} funding from the European Research Council (ERC), project DOC (The Dawn of Organic Chemistry), contract 741002 and \textit{(iv)} supported by the project PRIN-INAF 2016 The Cradle of Life - GENESIS-SKA (General Conditions in Early Planetary Systems for the rise of life with SKA). {D.S. acknowledges support from the Heidelberg Institute of Theoretical Studies for the project "Chemical kinetics models and visualization tools: Bridging biology and astronomy".}

%=============================================
%
%-----------------------------------------------------------------
%------- BIBLIO -------------------
%-----------------------------------------------------------------
%
%\bibliographystyle{aasjournal}
%\bibliography{/Users/cecilefavre/Documents/articles/biblio}

%%%%%%%%%%%%%%%%%%%%%%%%%%%%%%%%%%%%%%%%%%%%%%%%%%%%%%%%%%%%%%%%%%%%%%%%%%%%%%%%

%===============================================================
%===============================================================

\end{document}